\documentclass[11pt]{article}
\usepackage{graphicx, amsmath, amsthm, amssymb,float}

\usepackage{natbib}

\def\e{\varepsilon}
\def\t{\theta}
\def\s{\sigma}
\def\b{\beta}
\def\mA{\mathcal{A}}
\def\mR{\mathcal{R}}
\def\hth{\hat{\theta}}
\def\hs{\hat{\sigma}}
\def\ts{\tilde{\s}}
\def\hW{\hat{W}}
\def\hH{\hat{H}}

\def\hb{\hat{\beta}}
\title{The Residual Information Criterion, Corrected}
\author{Chenlei Leng
  \footnote{Leng is Assistant Professor,
    Department of Statistics and Applied Probability, 
    National University of Singapore. 
    Leng's research is supported in part by NUS research grant 
    R-155-050-053-133 (Email: stalc@nus.edu.sg). } 
}
\date{\today}
\begin{document}
\maketitle
  %=======
\begin{abstract}
  Shi and Tsai (JRSSB, 2002) 
  proposed an interesting residual information criterion
  (RIC) for model selection in regression. 
  Their RIC was motivated by the principle 
  of minimizing the Kullback-Leibler discrepancy between the residual
  likelihoods of the true and candidate model. We show, however, under 
  this principle, RIC would always choose the full (saturated) model. 
  The residual likelihood therefore, is not appropriate as a discrepancy 
  measure in defining information criterion.
  We explain why it is so and
  provide a corrected residual information criterion as a remedy.
  \\
  \\
  \noindent
      {\bf KEY WORDS:} 
      {\it 
	Residual information criterion;
	Corrected residual information criterion. }
\end{abstract}
  
%======================  
\section{Introduction}
Given $n$ iid observations from a true model
\[y =X \beta_0+\e, \]
where $y=(y_1,...,y_n)'$, $X$ is a $n \times p$ design matrix,
$\e=(\e_1,...,\e_n)'$ follows a multivariate distribution
with mean 0 and variance $\s_0^2 W(\t_0)$, and $\b_0 \in \mR^{p \times 1}$ 
is an unknown vector to be estimated. Here $\t_0$ is an $m \times 1$ vector
parameterizing the correlation matrix. Finally, we denote
$\mA_0=\mA(\b_0)=\{j:\beta_{0j} \ne 0,~j=1,...,p\}$ as the nonzero
coefficient set and $k_0=\# \mA_0$ as the number of nonzero coefficients.  
The problem of estimating $\mA_0$ is often 
referred to as variable selection or model selection.

Variable selection in linear regression is probably one
of the most important problems in statistics. See for example the references
in Shao (1997). To automate the process
of choosing a finite dimensional candidate model out of all possible models,
various information criteria have been developed. There are two basic
elements in all of these criteria: 
one element that measures the goodness of fit and the other term which
penalizes the complexity of the fitted model, usually taken
as a function of the parameters used. Generally speaking, the
existing variable selection approaches can be classified into two broad 
categories. On one hand,
AIC type of criteria, such as AIC (Akaike, 1970) and AICc
(Hurvich and Tsai, 1989), seek to minimize the
Kullback-Leibler divergence between the true and candidate model.
On the other hand, BIC (Schwarz, 1978) type of criteria are used to identify
a candidate model to achieve selection consistency. 
Obviously, these criteria are motivated by different assumptions and
different considerations, practically and theoretically. Any particular
choice on which one to use probably depends on the context and is subject to 
criticism, as each has its own merits and shortcomings. 

In an important paper, Shi and Tsai (2002) proposed an interesting information
criterion termed the residual information criterion (RIC). 
The authors showed that RIC is motivated by the consideration of minimizing
the discrepancy between the residual log-likelihood functions of
the true and candidate model. However, surprisingly, the authors 
arrived at a BIC type of criterion, in marked contrast with some other
information criteria, such as AIC, AICc, motivated by the same principle
of minimizing Kullback-Leibler discrepancy.

In this paper, we show that the RIC approach is not targeting at 
minimizing the Kullback-Leibler discrepancy between residual likelihoods. 
We provide a corrected 
criterion $\text{RIC}^*$ motivated by this principle.
However, we show that if the residual likelihoods are used to evaluate the 
Kullback-Leibler divergence between models, 
RIC (i.e. $\text{RIC}^*$) would always choose the full model.
Therefore, the residual likelihood is not an
appropriate loss function to define an information criterion. We
provide a simple likelihood based approach to circumvent the problem.

The rest of the paper is organized as follows. Section 2 reviews the RIC
method in Shi and Tsai. Since Shi and Tsai's RIC is not approximating 
the Kullback-Leibler divergence, we provide the $\text{RIC}^*$ measure as
a correction. However, $\text{RIC}^*$ always chooses the
full model and the reason is explained.
Section 3 presents the correct residual likelihood information criterion,
motivated by minimizing the Kullback-Leibler divergence between likelihoods
instead of residual likelihoods. Concluding remarks are given in Section 4.

%=============
\section{The Residual Information Criterion}
We review the RIC method in Shi and Tsai (2002) in this section.
The model we consider in this article is a special case of that in
Shi and Tsai (2002) by assuming the Box-Cox transformation parameter $\lambda$
is 1. The results in the paper can be easily extended to Box-Cox models
following similar arguments in Shi and Tsai.

We start by looking at a candidate (working) model
\[y=X\b+\e, \]
such that $\# \mA(\b)=k$. We denote the active covariates in $X$ as $X_\mA$.
Inspired by the residual likelihood method in Harville (1974) or 
Diggle {\it et al.} (1994) to obtain unbiased estimator for the
error variance, we can write the residual log-likelihood
as 
\begin{align}
 L(\theta',\sigma^2)=&-\frac{1}{2}(n-k)\log(2\pi)+\frac{1}{2}
 \log|X_\mA'X_\mA|-\frac{1}{2}(n-k)\log(\s^2)-\frac{1}{2} \log|W| \notag\\ 
& -\frac{1}{2}\log| X_\mA'W^{-1}X_\mA|-\frac{1}{2}
y'(W^{-1}-H_\mA)y/\s^2, \label{eq:rl}
\end{align}
where $H_\mA = W^{-1}X_\mA'(X_\mA'W^{-1}X_\mA)^{-1}X_\mA'W^{-1}$ 
and the dependence of $W$
on $\t$ is suppressed. A useful measure of the distance between the working
model and the true model is the Kullback-Leibler divergence
\begin{equation}
  d(\t',\s^2)=E_0[-2L(\t',\s^2)+2L_0(\t_0',\s^2_0) ], \label{eq:kl}
\end{equation}
where $E_0$ denotes the expectation under the true model and $L_0$ denotes
the residual log-likelihood of the true model. Clearly, the best model loses
the least information, in terms of Kullback-Leibler distance,
relative to the truth and is therefore preferred. Such a criterion formulates
RIC in an information-theoretical framework. Provided that one can unbiasedly
estimate $d(\t',\s^2)$, this criterion provides sound basis for parameter 
estimation and statistical inference under appropriate conditions.

Since $E_0[2L_0(\t_0',\s^2_0)]$ is
independent of the working model, we just need to evaluate
$E_0[-2L(\t',\s^2)]$. In Shi and Tsai (2002), (\ref{eq:kl}) is
written as
\begin{align}
 d(\t',\s^2)  = E_0\Big[(n-k)\log(\s^2)+\log|W|&+\log| X_\mA'W^{-1}X_\mA|
   \notag \\
  & +y'(W^{-1}-H_\mA)y/\s^2 \Big] \label{eq:kl0}\\
   =(n-k)\log(\s^2)+\log|W|&+\log| X_\mA'W^{-1}X_\mA|\notag\\
   &+E_0(X\b_0+\e)'(W^{-1}-H_\mA)(X\b_0+\e)/\s^2 \label{eq:kl1}
\end{align}
by omitting irrelevant terms. 
By substituting their estimated values
$\hat{\t}$, $\hat{\s}^2$ into (\ref{eq:kl1}), we have
\begin{align}
 d(\hth',\hs^2)  = (n-k)\log(\hs^2)&+\log|\hW|+\log| X_\mA'\hW^{-1}X_\mA|
   \notag \\
   &+(X\b_0)'(\hW^{-1}-\hH_\mA)(X\b_0)/\hs^2+ 
   \text{tr}\{ (\hW^{-1}-\hH_\mA)W_0\} \s_0^2/\hs^2.
   \label{eq:kl2}
\end{align}
The above expression involves an unknown quantity $\s_0^2$. Following 
Shi and Tsai, we judge
the quality of the candidate model by $E_0\{ d(\hth',\hs^2)\}$.
Now, if we assume $\mA_0 \subseteq \mA$, an assumption also used in deriving
AICc (Hurvich and Tsai, 1989),
the third term becomes zero. Furthermore,
if we assume $\hth$ is consistent for $\t_0$, we can estimate $W_0$ by $\hW$
since $\hW=W_0+o_p(1)$. Then the fourth term can be approximated as
$(n-k)\s_0^2/\hs^2$. Since  $\mA \subseteq \mA_0$, $(n-k)\hs^2/\s_0^2$ then
follows $\chi^2_{n-k}$ distribution and therefore
\[E_0[(n-k)\s_0^2/\hs^2]=(n-k)^2/(n-k-2).\] 
Finally, Shi and Tsai argued that $\log| X_\mA'\hW^{-1}X_\mA|$ can be 
approximated by $k\log(n)$. Putting everything together, they proposed the
residual information criterion as follows
\begin{equation}
  \text{RIC}= (n-k)\log(\hs^2)+\log|\hW|+k\log(n)-k+\frac{4}{n-k-2}, 
  \label{eq:ric}
\end{equation}
after removing the constant $n+2$. Asymptotically,
the complexity part of RIC is of the order $k\log(n)$.  Comparing to 
$\text{BIC} = n\log(\ts^2)+k\log(n)$, where $\ts^2$ is the MLE of $\s_0^2$,
it is intuitively clear that Shi and Tsai's RIC yields
consistent models as BIC does. The complexity penalty of RIC, however, is 
fundamentally different
from that of other familiar information criterion such as AIC and AICc,
designed to approximate the Kullback-Leibler divergence between two models. 
This observation raises the question
on whether RIC rightfully approximates the divergence. 

It turns out that Shi and Tsai's derivation 
motivated by minimizing the Kullback-Leibler distance, is 
incorrect in at least two important places:
\begin{itemize}
  \item[1.] In (\ref{eq:kl0}), a model dependent term $\log|X_\mA'X_\mA|$ 
    is omitted from (\ref{eq:rl}), which causes serious bias in deriving
    an information criterion. In fact, following Shi and Tsai's arguments,
    we can approximate $\log| X_\mA'X_\mA|$ by $k\log(n)$ and thus, RIC should
    have been
    \[\text{RIC}^*= (n-k)\log(\hs^2)+\log|\hW|-k+\frac{4}{(n-k-2)}.\]
    Note that in this formulation, $\text{RIC}^*$ always chooses the full 
    model.
  \item[2.] Even more severely, 
    the practice of approximating the Kullback-Leibler distance
    between residual likelihoods for comparing models
    is totally wrong. To illustrate, suppose that $W=I$. In this simple
    case, the residual likelihood becomes
    \[ L(\sigma^2)=
    -\frac{1}{2}(n-k) \log(\sigma^2)-\frac{1}{2}y'
    [I-X_\mA(X_\mA'X_\mA)^{-1}X_\mA]y/\s^2.\]
    We see immediately that 
    $E_0[-2L(\sigma^2)]=(n-k)\log(\s^2)+(n-k)\s_0^2/\s^2$
    whenever $\mA_0 \subseteq \mA$. Thus, for candidate models that include
    $X_{\mA_0}$ in the covariate set,
    $E_0[-2L(\sigma^2)]$ is always minimized by $\s^2=\s_0^2$
    and in this case $E_0[-2L(\sigma^2)]=(n-k)(\log(\s^2_0)+1)$. Therefore,
    if one knows the exact data generating process, the ideal RIC leads to
    the full model, as its $E_0[-2L(\sigma^2)]$ is the smallest. This explains
    why $\text{RIC}^*$ always chooses the full model. 
\end{itemize}
Given the above serious flaws in going from deriving unbiased estimator of the
Kullback-Leibler divergence to RIC, Shi and Tsai's RIC in (\ref{eq:ric})
seems improperly motivated.
Fortunately, Shi and Tsai's derivation can be corrected and
we introduce a corrected RIC in the next section.

%=====================
\section{A Corrected Residual Information Criterion}
Instead of using the 
residual likelihood, a justifiable criterion is to use the log-likelihood
\[L(\b', \t', \s^2)=n\log(\s^2)+\log|W|+(y-X\b)'W^{-1}(y-X\b) \]
in defining the divergence
\[d(\b',\t',\s^2)=E_0[-2L(\b',\t',\s^2)+2L_0(\b_0', \t_0',\s^2_0)].\]
We can write
\begin{align*}
  E_0[-2L(\b',\t',\s^2)]&=E_0 \big[n \log(\s^2)+\log|W|+(X\b_0+\e-X\b)'W^{-1}
  (X\b_0+\e-X\b) \big]\\
  &=n \log(\s^2)+\log|W|+n\s_0^2/\s^2+(X\b-X\b_0)'W^{-1}(X\b-X\b_0)\s_0^2/\s^2.
\end{align*}
We can now replace $\s^2$, $\b$ and $\t$ by the their estimates by using
the residual likelihood method. Now, suppose that $\mA_0 \subseteq \mA$.
Following Shi and Tsai again,
$E_0 n\s_0^2/\hs^2 \approx n(n-k)/(n-k-2)$. Since 
$\hb-\b_0$ follows normal distribution 
$N\{0, \s_0^2 (X_\mA'W^{-1}X_\mA)^{-1}\}$
asymptotically, 
\[\frac{1}{k}(X\hb-X\b_0)'W^{-1}(X\hb-X\b_0)\s_0^2/\hs^2 \]
is distributed approximately as $F(k, n-k)$. Therefore,
\[E_0\{ (X\hb-X\b_0)'W^{-1}(X\hb-X\b_0)\s_0^2/\hs^2\} =\frac{k(n-k)}{n-k-2}.\] 
Putting everything together, we have the following corrected
residual information criterion, which we shall refer to
as RICc,
\[ \text{RICc}=n \log(\hs^2)+k+\frac{4(k+1)}{n-k-2},  \]
by omitting a constant $n+2$.
Note that 
\[\text{AIC} = n\log(\ts^2)+2k, \]
and 
\[ \text{AICc} = n\log(\ts^2)+2n(k+1)/(n-k-2) \]
where $\ts^2$ is the MLE of $\s_0^2$. We can decompose the first expression
of RICc, AIC and AICc as $n\log(\text{RSS})-n\log(n-k)$,
$n\log(\text{RSS})-n\log(n)$ and $n\log(\text{RSS})-n\log(n)$
respectively. Thus, the complexity penalties for RICc, AIC, AICc are
$-n\log(n-k)+k+4(k+1)/{(n-k-2)}$, $-n\log(n)+2k$ and
$-n\log(n)+2n(k+1)/(n-k-2)$
respectively. It can
be seen that RICc has a larger penalty function than AIC 
and a smaller penalty than AICc when $n \gg k$.
%======================
\section{Concluding Remarks}
In fitting a model to data, one is required to
choose a set of candidate models, 
a fitting procedure and a criterion to compare competing models. A minimal
requirement for a reasonable criterion is that the population version of the
criterion is uniquely minimized by the set of the parameters which 
generate the data. The population version of the residual likelihood 
information criterion is minimized by the full model and thus fails to meet 
this basic requirement. Therefore, the residual likelihood cannot be used as
a discrepancy measure between models. A simple remedy is to use the
likelihood based Kullback-Leibler divergence. 

Being a legitimate criterion on its own,
our arguments show that Shi and Tsai's RIC is not motivated by the
right principle. Should one have followed their motivation, 
RIC (i.e. $\text{RIC}^*$ by our notation) 
would have always chosen the full model. 
However, Shi and Tsai's RIC, though motivated by the wrong principle (using the
residual likelihood instead of the likelihood) and ignoring dangerously
an important term $\log|X'X|$ in approximation, has good small sample 
performance in their 
simulations. Additionally, Shi and Tsai's RIC has been successfully
applied to a number of applications, such as normal linear regression, Box-Cox
transformation, inverse regression models (Ni {\it et al.}, 2005) and 
longitudinal data analysis (Li {\it et al.}, 2006). The success may be 
understood as Shi and Tsai's RIC resembles BIC. Despite the increasing
popularity of RIC, Shi and Tsai's RIC remains unmotivated. It remains to find a
justification for Shi and Tsai's RIC as a future research topic.

\bibliographystyle{apalike}

%\bibliography{paper}

\begin{thebibliography}{}

\bibitem[Akaike, 1970]{Akaike:1970}
  Akaike, H. (1970).
  \newblock Statistical predictor identification.
  \newblock {\em Annals of Institute of Statistical Mathematics}, 22, 203-217.

\bibitem[Azari, et al 2006]{Azar:etal:2006}
  Azari, R., Li, L., and Tsai, C.-L. (2006).
  \newblock Longitudinal data model selection.  
  \newblock {\em Computational Statistics and Data Analysis}, 50, 3053-3066. 

\bibitem[Diggle, et al. 1994]{Diggle:etal:1994}
  Diggle, P.J., Heagerty, P.J., Liang, K.-Y. and Zeger, S.L. (2002).
  \newblock Analysis of longitudinal data. (2nd edition).
  \newblock Oxford: Oxford University Press.

\bibitem[Harville, 1974]{Harv:1974}
  Harville, D.A. (1974).
  \newblock Bayesian inference for variance components using only error 
  contrasts.
  \newblock {\em Biometrika}, 61, 383-385.

\bibitem[Hurvich and Tsai, 1989]{Hurv:Tsai:1989}
  Hurvich, C. M., and Tsai, C.-L. (1989). 
  \newblock Regression and time series model selection in small samples. 
  \newblock {\em Biometrika}, 76, 297-307.

%\bibitem[Mallow, 1973]{Mallow:1973}
%  Mallow, C.L. (1973).
%  \newblock Some comments on Cp.
%  \newblock {\em Technometrics}, 15, 661-675.

\bibitem[Ni, et. al, 2005]{Ni:etal:2005}
  Ni, L., Cook, R. D., and Tsai, C-L. (2005).
  \newblock A note on shrinkage sliced inverse regression.
  \newblock {\em Biometrika}, 92, 242-247.

\bibitem[Schwarz, 1978]{Schw:1978}
  Scharwz, G. (1978).
  \newblock Estimating the dimension of a model.
  \newblock {\em Annals of Statistics}, 6, 461-464.

\bibitem[Shao, 1997]{Shao:1997}
  Shao, J. (1997). 
  \newblock An asymptotic theory for linear model selection (with discussion).
  \newblock {\em Statistica Sinica}, 7, 221-264.

\bibitem[Shi and Tsai, 2002]{Shi:Tsai:2002}
  Shi, P. and Tsai, C.-L. (2002).
  \newblock Regression model selection-a residual likelihood approach.
  \newblock {\em Journal of the Royal Statistical Society B}, 64, 
   237--252.
\end{thebibliography}
\end{document}